# Misaligned spin and orbital axes cause the anomalous precession of DI Herculis[*]

Simon Albrecht[†,‡], Sabine Reffert[§], Ignas A.G. Snellen[†], & Joshua N. Winn[‡]

**The orbits of binary stars precess as a result of general relativistic effects, forces arising from the asphericity of the stars, and forces from additional stars or planets in the system. For most binaries, the theoretical and observed precession rates are in agreement[1]. One system, however—DI Herculis—has resisted explanation for 30 years[2–4]. The observed precession rate is a factor of four slower than the theoretical rate, a disagreement that once was interpreted as evidence for a failure of general relativity[5]. Among the contemporary explanations are the existence of a circumbinary planet[6] and a large tilt of the stellar spin axes with respect to the orbit[7,8]. Here we report that both stars of DI Herculis rotate with their spin axes nearly perpendicular to the orbital axis (contrary to the usual assumption for close binary stars). The rotationally induced stellar oblateness causes precession in the direction opposite to that of relativistic precession, thereby reconciling the theoretical and observed rates.**

We observed DI Herculis (HD 175227, apparent $V$ magnitude outside eclipse = 8.45, spectral types B4V and B5V, orbital period $P$ = 10.55 d, orbital eccentricity $e$ = 0.49)[9] with Sophie[10], a high resolution échelle spectrograph on the 1.93m telescope of the Observatoire de Haute-Provence. Fig. 1 shows four representative spectra obtained during the primary eclipse and four from the secondary eclipse. Supplementary Figs. 1–3 show all the spectra.

By analysing the distortions of the spectral lines that occur during eclipses (the "Rossiter-McLaughlin effect"[11,12,13]) the relative orientation of the spin and orbital axes as projected on the sky can be derived. Our model of the spectra and of the RM effect is based on the method of ref. 14, as summarised here. Because of the light from the foreground star, we could not treat the RM effect as a simple wavelength shift, as is commonly done for eclipses of stars by planets[15,16]. Instead we modelled the line profiles of both stars, taking into account stellar rotation, surface velocity fields, orbital motion, and partial blockage during eclipses. By adjusting the parameters of the model to fit the observed spectra, we derived estimates of the orbital and stellar parameters, including the angles $\beta_p$ and $\beta_s$ (subscripts 'p' and 's' refer to primary and secondary) between the sky projections of the spin axes and the orbital axis. The angles are defined such that $\beta = 0°$ when the axes are parallel, and $\beta = \pm 90°$ when they are perpendicular.

---

[†] Leiden Observatory, Leiden University, Postbus 9513, 2300 RA Leiden, The Netherlands
[‡] Department of Physics, and Kavli Institute for Astrophysics and Space Research, Massachusetts Institute of Technology, Cambridge, MA 02139, USA
[§] Zentrum für Astronomie Heidelberg, Landessternwarte, Königstuhl 12, 69117, Heidelberg, Germany

We focused on the Mg II line at 4,481Å because it is the strongest line that is broadened mainly by stellar rotation. However, it is located in the red wing of the stronger pressure-broadened He I line at 4,471Å. Therefore, we fitted a Lorentzian model to the encroaching wing of the He I line and subtracted it before modelling the Mg II line. (Similar results were obtained when both lines were modelled simultaneously.)

In the model, the simulated absorption lines were shifted in wavelength according to the Keplerian radial velocity, and were broadened by anisotropic macroturbulence (using the parameterization of Gray[17]) and rotation (parameterized by $v_p \sin i_p$ and $v_s \sin i_s$). For each star, and for each phase of the eclipse, we modelled a limb-darkened, uniformly rotating stellar disk with several thousand pixels. Each pixel was further divided into 17 sub-pixels to represent the distribution of turbulent velocities. The light from the sub-pixels was Doppler shifted according to a Gaussian distribution for the tangential and radial turbulent velocity fields ($\xi_{R,T}$), assuming the tangential and radial fields to have equal amplitudes and equal surface fractions. (Although we included macroturbulence for completeness, the results for $\beta_p$ and $\beta_s$ do not depend on the details, and similar results were obtained even when macroturbulence was ignored.)

It was also necessary to include parameters for the equivalent widths of the Mg II lines (Table 1). Finally, we adopted values of the stellar radii from the literature (Table 1), and assumed the limb-darkening law to be linear with a coefficient $u = 0.4$ at 4,500 Å for both stars[17]. The emergent spectra from the uneclipsed pixels of both stars were summed to create the model spectrum.

We used Levenberg-Marquardt least-squares minimization[18] to derive the best fitting parameters, and the bootstrap method[19] to derive the statistical 1-$\sigma$ parameter uncertainties (Tables 1 and 2). This uncertainty estimate did not take into account some possible systematic errors. We investigated the possible error due to gravity brightening by assuming the emergent intensity to be proportional to the local acceleration due to gravity[20], finding the resultant changes in $\beta_p$ and $\beta_s$ to be smaller than 1° and therefore negligible. Likewise, we did not allow for the possible effects of differential rotation, nor did we allow for any changes in the limb-darkening law within the lines as compared to the continuum. We found no evidence for those effects, though it remains possible that they are present and that our macroturbulence model is compensating for them. In addition, because the lines are relatively broad, errors in the normalization of the continuum could influence the results. To account for these effects in our error analysis, we re-fitted our model to a collection of spectra normalized using different functional forms, and choosing different wavelength domains over which those functions are fitted. We also re-fitted using limb-darkening coefficients ranging from 0 to 1, and characteristic macroturbulent velocities in the range 5–50 km s$^{-1}$. From the scatter in the results for the fitted parameters, we derived our final 1-$\sigma$ uncertainties in the $v \sin i$ and $\beta$ parameters, with the largest effects arising from changes in normalization.

Figure 1 shows the observed spectra after subtracting the model spectrum of the foreground star, for clarity. This figure also shows the best-fitting model, and the model assuming co-aligned spins and orbit, which fits the data poorly. Our main result is that the stellar rotation axes are strongly tilted with respect to each other, and with respect to the projected orbital axis: $\beta_p = 72 \pm 4°$ and $\beta_s = -84 \pm 8°$. Although we reiterate that our quantitative results were obtained by fitting the line profiles of both stars directly, in Fig. 2 we show the apparent radial velocity of each star, a form of presentation that may make the result of a large misalignment more obvious.

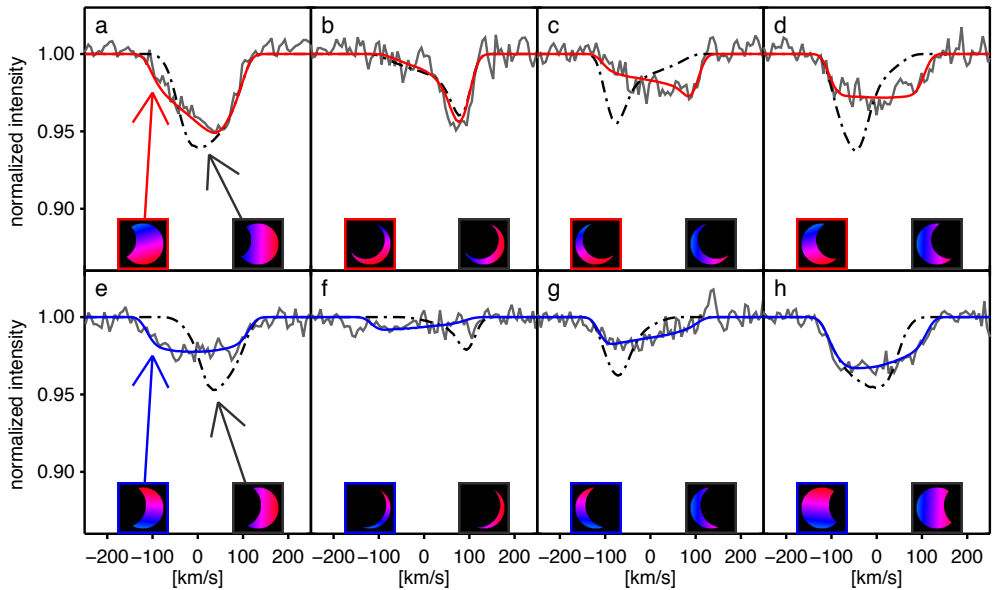

**Figure 1 | Distortion of stellar absorption lines during eclipses**. We observed DI Herculis during eclipses on 20/21 May, 13/14 July, and 30 June / 1 July 2008. We obtained 22 spectra during primary eclipses, 15 spectra during the secondary eclipse and 15 spectra outside of eclipses. During an eclipse the lines of the background star are distorted because they are missing some of the components that produce rotational broadening. This phenomenon, the Rossiter-McLaughlin effect, was foreseen as early as 1893 (ref. 11) and observed definitively in 1924 (refs. 12, 13). We began with the reduced two-dimensional spectra delivered by the Sophie instrument software. Bad pixels were identified and removed, and the continuum level in each order was normalized. The uncertainty in the wavelength scale is equivalent to a few meters per second, and is negligible for our purposes. The spectra have a resolution of 40,000 and the median signal-to-noise ratio is 80 in the vicinity of 4,500 Å. Each panel shows the Mg II line (4,481 Å) of the eclipsed star. For clarity, a model of the lines of the foreground star was subtracted, and the data were binned by four pixels. The upper panels (**a**-**d**) show data from the primary eclipse, and the lower panels (**e**-**h**) show data from the secondary eclipse. (Supplementary Figs. 1-3 show all the spectra.) The gray solid lines are the data, and the red and blue lines are the best-fitting model, with $\beta_p = 72°$ and $\beta_s = -84°$. The dashed-dotted lines show the model assuming co-aligned stellar and orbital axes, which gives a poor fit to the data. Each panel has two inset illustrations. The left inset illustrates the line-of-sight component of the rotational velocity of the eclipsed star in the best-fitting model, with blue and red indicating approaching and receding velocities, respectively. The right insets show the same for the co-aligned model.

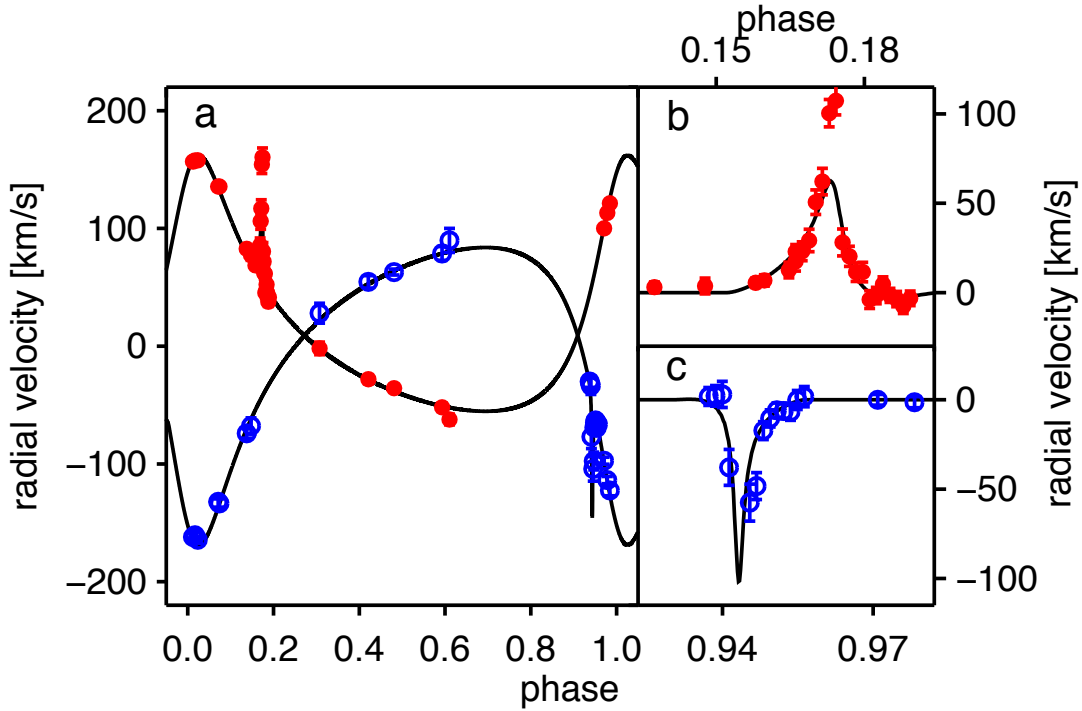

**Figure 2 | Apparent radial velocities of DI Herculis. a**, The apparent radial velocity (RV) of the primary (red filled circles) and secondary (blue open circles) as a function of orbital phase. The solid line is the calculated radial velocity based on our model, including Keplerian motion and the Rossiter-McLaughlin effect. Outside of eclipses, the line positions were determined by fitting Gaussian functions. During eclipses, the apparent radial velocity was defined as the intensity-weighted mean wavelength of the line. **b,c,** Close-ups of the Rossiter-McLaughlin anomaly during the primary eclipse (upper) and secondary eclipse (lower). The anomaly is a redshift throughout the primary eclipse, and a blueshift throughout the secondary eclipse, indicating large spin-orbit misalignments. Had the spins and orbit been aligned, the anomaly would have been a redshift during the first half of the eclipse, and a blueshift during the second half. The error bars shown are 1-σ errors based only on the signal-to-noise ratio of the spectra. For radial velocities obtained during eclipses, the true errors are larger due to imperfect subtraction of the foreground light, which is why there are apparently large deviations between the data and the model at mid-eclipse (indeed, one radial velocity at the secondary mid-eclipse lies outside the plotted range). We emphasize that the eclipse radial velocities are shown here for illustration only, as a concise visual summary of the complex distortions of the eclipse spectra: they were not used in our quantitative calculations.

The factor-of-four discrepancy between the theoretical and observed rates of apsidal precession was based on the premise that the stellar spin axes are perpendicular to the orbital plane. Now that this premise has been proved false, the theoretical precession rate must be recalculated. For this calculation it is necessary to know the true angles between the spin and orbital axes, and the actual rotation velocities of the stars, even though only their sky projections are measured. The unmeasured angles are the inclinations $i_p$ and $i_s$ of the stellar spin axes with respect to the line of sight. We performed a Monte Carlo calculation of the theoretical precession rate[7,8], in which the cosine of each inclination was drawn from a uniform distribution. We excluded inclinations giving rotation speeds faster then 600 km s$^{-1}$, the approximate breakup speeds. To calculate the precession due to the rotationally induced oblateness, we adopted apsidal motion constants of $k_{2p} = 0.0087$ and $k_{2s} = 0.0081^4$. The oblateness causes *retrograde* apsidal motion (–2.23 arcsec/cycle), while tidal deformations and general relativity cause *prograde* apsidal motion (+1.35 and +2.40 arcsec/cycle, respectively). The net theoretical precession rate is $1.52^{+0.06}_{-5.4}$ arcsec/cycle, where the value quoted is the mode of the Monte Carlo distribution, and the quoted error interval encloses 68.3% of the results. This agrees with the observed precession rate of 1.08 ± 0.16 arcsec/cycle (ref. 21 and Fig. 3). We therefore consider the mystery of the anomalous apsidal precession of DI Herculis to be solved.

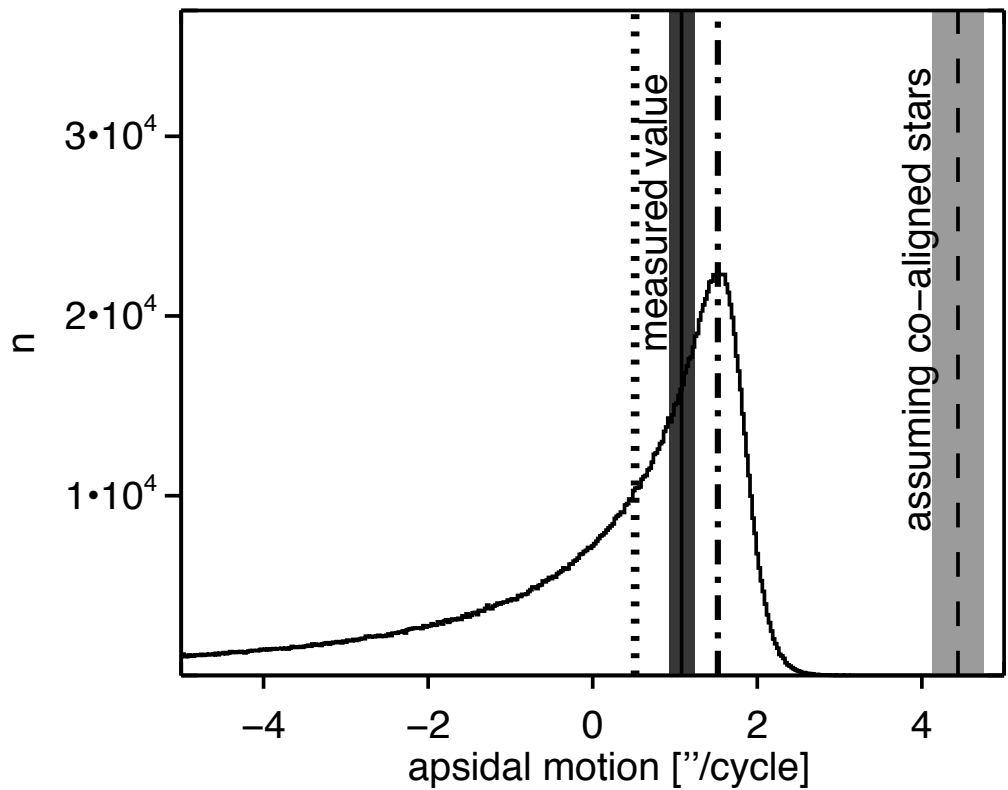

**Figure 3 | Apsidal motion of DI Herculis, calculated and observed**. The plot shows the observed apsidal motion[21] of 1.08 ± 0.16 arcsec per cycle (solid line and dark gray area), the apsidal motion expected assuming aligned axes[3] of 4.44 ± 0.31 arcsec per cycle (dashed line and light gray area on the right side of the plot), and the histogram of the expected apsidal motion based on our results (solid black line). For the expected apsidal motion, we performed a Monte Carlo calculation in which stellar inclinations are drawn from a uniform distribution in cos *i* after restricting the stellar rotation speed to be slower than 600 km s$^{-1}$. The median of the distribution (0.52 arcsec per cycle) is indicated by the dotted line and the mode by the dashed-dotted line (1.52 arcsec per cycle). The long tail to negative values corresponds to rapid rotation in a nearly pole-on configuration.

The solution to the mystery of DI Herculis raises the questions of how the stars became so strongly tilted with respect to their orbit. It seems unlikely that they formed in such a state, although it is not clear at what point in their relatively short lifetime (~5 Myr[4]) the misalignment occurred. One possible mechanism is the Kozai effect, whereby a third body in the system excites oscillations of the binary's eccentricity and orbital orientation[22]. Whatever the route to the present configuration of DI Herculis, it would be interesting to know whether it is unique or whether spin-orbit misalignment is more common among close binaries than has been expected. A survey to measure the relative orientations of spin axes in selected eclipsing binaries, or systems that can be resolved with interferometry[23,24], may help to answer these questions.

**Acknowledgements** We are grateful to Á. Giménez, E. Guinan, and T. Mazeh for bringing DI Herculis to our attention. We thank J. Lub, R. Tubbs, C. Hopman, Y. Levin, and D. Fabrycky for discussions about double stars and their properties. We also thank H. Reckman for essential help during one observing run. We are grateful to the Sophie team for building the spectrograph and reduction pipeline. This research has made use of the SIMBAD database, operated at CDS, Strasbourg, France and the Vienna Atomic Line database (VALD) located at http://ams.astro.univie.ac.at/vald/. S. Albrecht acknowledges support during part of this project by a Rubicon fellowship from the Netherlands Organisation for Scientific Research (NWO). J.N.W. acknowledges support from NASA Origins grant NNX09AD36G. S.A. and S.R. acknowledge funding from the Optical Infrared Coordination network (OPTICON).

**Author Contributions** S.A. participated in the development of the concept of this research and the analysis code, participated in the observations, the analysis and interpretation of the data and writing the manuscript. S.R participated in the development of the concept of this research and the analysis code and the observations. I.A.G.S. participated in the analysis and interpretation of the data and in writing the manuscript. J.N.W. participated in the analysis and interpretation of the data and in writing the manuscript.

**Author Information** Correspondence and requests for materials should be addressed to S.A. (albrecht@space.mit.edu).

**References cited**

1. Claret, A. & Giménez, Á. The Apsidal Motion Test of the Internal Stellar Structure - Comparison Between Theory and Observations. *Astron. Astrophys*. **277**, 487-502 (1993)

2. Martynov, D. I. & Khaliullin, K. F. On the relativistic motion of the periastron in the eclipsing binary system DI Herculis *Astrophys. SpaceScience*, **71**, 147-170 (1980)

3. Guinan, E. F. & Maloney, F. P. The apsidal motion of the eccentric eclipsing binary DI Herculis - an apparent discrepancy with general relativity. *Astron. J.* **90**, 1519-1528 (1985)

4. Claret, A. Some notes on the relativistic apsidal motion of DI Herculis. *Astron. Astrophys*. **330**, 533-540 (1998)

5. Moffat, J. W. The orbital motion of DI Herculis as a test of a theory of gravitation. *Astrophys. J.* **287**, L77-L79 (1984)

6. Hsuan, K. & Mardling, R. A. A Three Body Solution for the DI Her System, *Astrophys. Space Science*, **304,** 243-246 (2006)

7. Shakura, N. I. On the Apsidal Motion in Binary Stars, *Soviet Astron. L.* 224-226 (1985)

8. Company, R., Portilla, M. & Giménez, Á. On the apsidal motion of DI Herculis. *Astrophys. J.* **335**, 962-964 (1988)

9. Popper, D. M. Rediscussion of eclipsing binaries. XIII - DI Herculis, a B-type system with an eccentric orbit. *Astrophys. J.* **254**, 203-213 (1982)

10. Perruchot, S. et al. The SOPHIE spectrograph: design and technical key-points for high throughput and high stability. *SPIE Conference Series*, **7014** (2008)

11. Holt, J. R. Spectroscopic Determination of Stellar rotation, *Astronomy and Astro-Physics*, **12**, 646 (1893)

12. Rossiter, R . A. On the detection of an effect of rotation during eclipse in the velocity of the brigher component of beta Lyrae, and on the constancy of velocity of this system. *Astrophys. J.* **60**, 15-21 (1924)

13. McLaughlin, D. B. Some results of a spectrographic study of the Algol system. *Astrophys. J.* **60**, 22-31 (1924)

14. Albrecht, S., Reffert, S., Snellen, I., Quirrenbach, A. & Mitchell, D. S. The spin axes orbital alignment of both stars within the eclipsing binary system V1143 Cyg using the Rossiter-McLaughlin effect. *Astron. Astrophys.* **474**, 565-573 **(**2007)

15. Queloz, D. et al. Detection of a spectroscopic transit by the planet orbiting the star HD209458. *Astron. Astrophys.* **359**, L13–L17 (2000)

16. Winn, J. N. et al. Measurement of Spin-Orbit Alignment in an Extrasolar Planetary System. *Astrophys. J.* **631**, 1215-1226, 2005

17. Gray, D.F., *The Observation and Analysis of Stellar Photospheres.* 3rd ed *Cambridge University Press,* ISBN 0521851866, (2005)

18. Markwardt, C. B. Non-linear Least Squares Fitting in IDL with MPFIT. Preprint at <http:arXiv.org/abs/0902.2850> (2009)

19. Press, W. H., Teukolsky, S. A., Vetterling, W. T., & Flannery, B. P., Numerical recipes in C. The art of scientific computing. 2nd ed., 689–699 *Cambridge: University Press*, ISBN 0521431085 (1992)

20. Von Zeipel H., The radiative equilibrium of a slightly oblate rotating star. *Mon. Not. R. Astron. Soc.*, **84**, 684-701(1924)

21. Guinan, E. F., Marshall, J. J. & Maloney, F. P. A New Apsidal Motion Determination for DI Herculis. *Information Bulletin on Variable Stars*, **4101**, 1. (1994)

22. Fabrycky, D. & Tremaine, S. Shrinking Binary and Planetary Orbits by Kozai Cycles with Tidal Friction. *Astrophys. J.* **669**, 1298-1315, (2007)

23. Chelli, A. & Petrov, R. G., Model fitting and error analysis for differential interferometry. II. Application to rotating stars and binary systems. *Astron. Astrophys. Suppl.* **109**, 401-415 (1995)

24. Le Bouquin, J.B. The spin-orbit alignment of the Fomalhaut planetary system probed by optical long baseline interferometry. *Astron. Astroph.* **498**, L14–L44 (2009)

25. Reisenberger, M. P. & Guinan, E. F. A possible rescue of general relativity in DI Herculis. *Astron. J*, **97**, 216-221 (1989)

26. Van Dien. E. Axial Rotation of the Brighter Stars in the Pleiades Cluster. *Journal of the Royal Astronomical Society of Canada*, **42**, 249-261 (1948)

**Table 1 | Orbital and physical parameters of DI Herculis**

| | Literature values | Ref. | This study |
|---|---|---|---|
| Orbital period, $P$ | 10.550164±0.000001 days | 3 | |
| Minimum light of primary eclipse, $T_{min}$ | 2442233.3481±0.0007 HJD | 3 | |
| Argument of periastron, $\omega$ | 330.2 ± 0.6° | 9 | |
| Orbital eccentricity, $e$ | 0.489 ± 0.003 | 9 | |
| Orbital inclination, $i_o$ | 89.3 ± 0.03° | 9 | |
| Primary velocity semiamplitude, $K_p$ | 110.7 ± 0.5 km s$^{-1}$ | 9 | 109 ± 1 km s$^{-1}$ |
| Secondary velocity semiamplitude, $K_s$ | 126.6±1.2 km s$^{-1}$ | 9 | 127 ± 1 km s$^{-1}$ |
| Primary additive velocity offset, $\gamma_p$ | 9.1 ± 0.3 km s$^{-1}$ | 9 | 7 ± 1 km s$^{-1}$ |
| Secondary additive velocity offset, $\gamma_s$ | = $\gamma_p$ | 9 | 11 ± 1 km s$^{-1}$ |
| Primary stellar radius, $R_p$ | 2.68 ± 0.05 $R_{sun}$ | 9 | |
| Secondary stellar radius, $R_s$ | 2.48 ± 0.05 $R_{sun}$ | 9 | |
| Primary stellar mass, $M_p$ | 5.15 ± 0.10 $M_{sun}$ | 9 | 5.1$^{+0.2}_{-0.1}$ $M_{sun}$ |
| Secondary stellar mass, $M_s$ | 4.52 ± 0.06 $M_{sun}$ | 9 | 4.4$^{+0.2}_{-0.1}$ $M_{sun}$ |
| Average equivalent width of Mg II | 0.27 Å | 9 | 0.24 Å |
| Primary macroturbulent velocity, $\xi_{R,T}$ | | | 12 ± 2 km s$^{-1}$ |
| Secondary macroturbulent velocity, $\xi_{R,T}$ | | | 10 ± 1 km s$^{-1}$ |
| Primary projected spin-orbit angle, $\beta_p$ | | | 72 ± 4° |
| Secondary projected spin-orbit angle, $\beta_s$ | | | –84 ± 8° |

HJD, heliocentric Julian day. The ref. 9 value for $\omega$ (329.9°), based on observations in 1968-1978, was precessed forward to 2008. We used two independent velocity offsets to account for effective centroid shifts that depend on spectral type, owing to temperature-dependent blended features. We allowed $K_p$, $K_s$, $\gamma_p$, and $\gamma_s$ to be free parameters. We fixed the values of $P$, $e$, $\omega$, $i_o$, and $T_{min}$ at values from the literature. If, instead, all the parameters are allowed to be free, we find results that are consistent with the literature values, but with reduced precision. We also find that the results for $\beta_p$ and $\beta_s$ agree with those obtained from the externally-constrained model.

**Table 2 | $v \sin i$ measurements**

| | 1948 (ref. 26) | 1974 (ref. 25) | 1985 (ref. 25) | 1988 (ref. 25) | 2008 (this study) |
|---|---|---|---|---|---|
| $v_p \sin i_p$ (km s$^{-1}$) | 76 | 34 ± 6 | 50 ± 30 | 61 ± 4 | 108 ± 4 |
| $v_s \sin i_s$ (km s$^{-1}$) | 76 | 50 ± 7 | 50 ± 30 | 75 ± 4 | 116 ± 4 |

We note that the apparent time variations in $v_p \sin i_p$ and $v_s \sin i_s$ were previously reported[25], and taken as possible evidence for spin precession (which would imply a substantial misalignment), although those authors could not draw a firm conclusion based on the available data. In principle, one might derive the three-dimensional orientation of the stellar spin axes by modelling the time variations in $v_p \sin i_p$ and $v_s \sin i_s$. At present, however, it is not possible to improve on the calculation of the theoretical precession rate using this technique, because the analysis hinges on the measurement from 1948 for which the uncertainty is unknown.

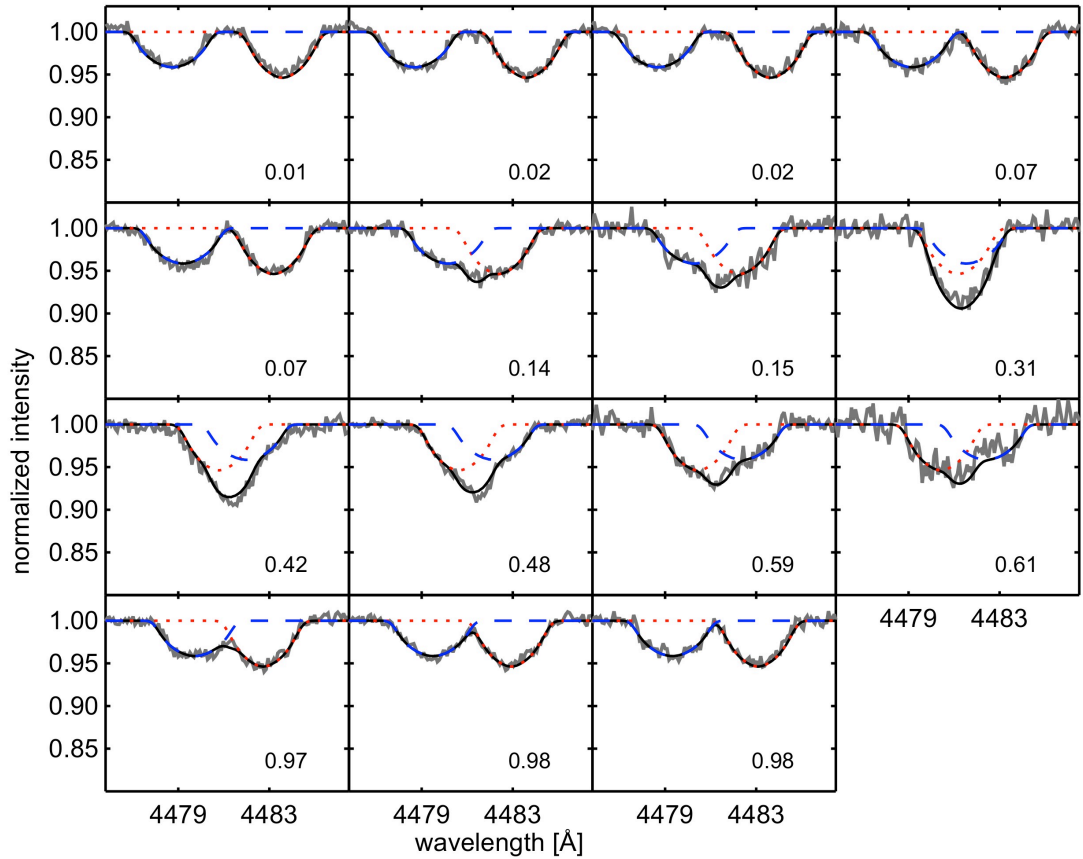

**Supplementary Figure 1 | Observations of DI Herculis outside of eclipses.** These panels show the out-of-eclipse spectra of both stars in the DI Herculis system, in the spectral region around the Mg II lines at 4,481 Å. The number in each panel indicates the orbital phase when the spectrum was taken. The gray solid lines represent the data, and the red dotted and blue dashed lines are the simulated absorption lines of the primary and secondary, respectively. The black line is the best-fitting model.

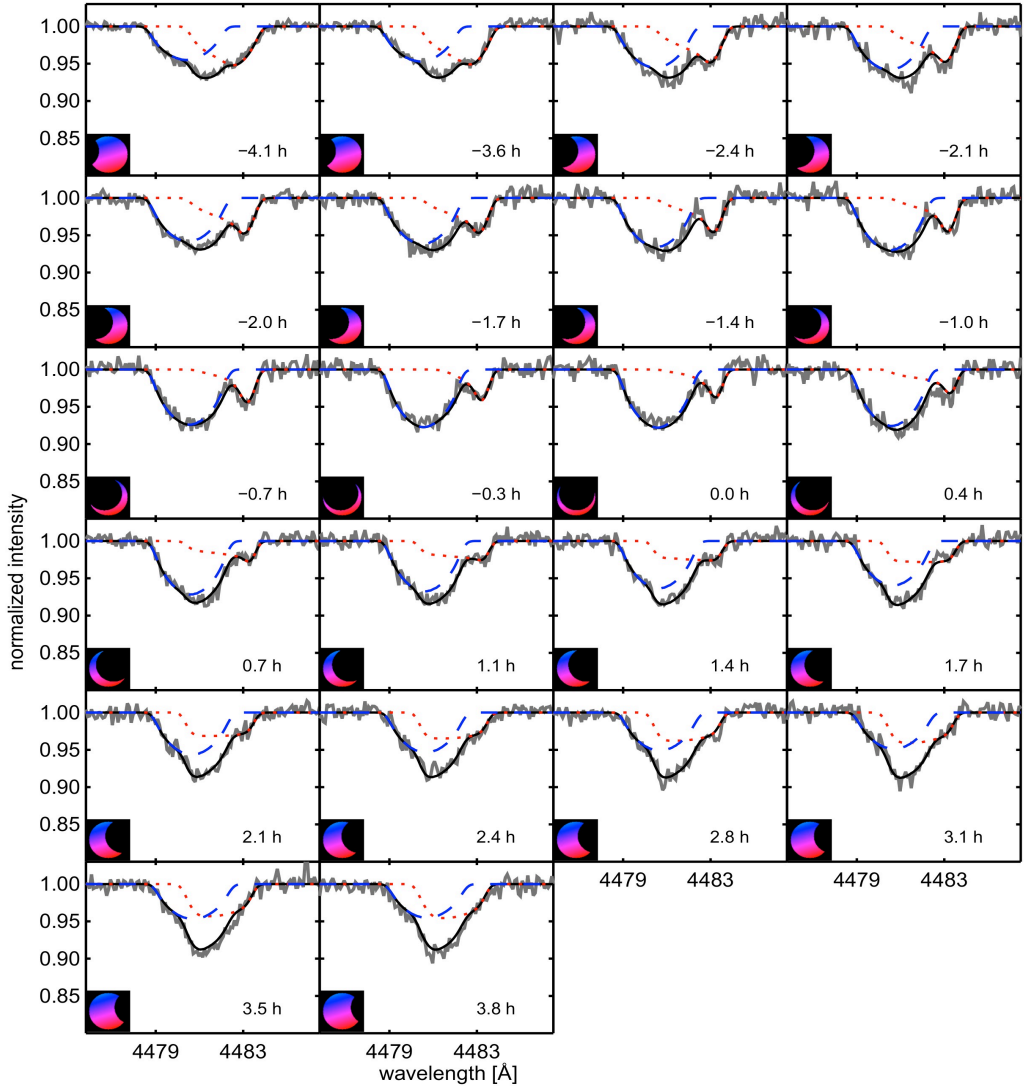

**Supplementary Figure 2 | Observations of DI Herculis during the primary eclipse.** These panels also show the spectra of both stars in the DI Herculis system, in the spectral region around the Mg II lines at 4,481 Å. However these spectra have been obtained during the primary eclipse. Here the number in each panel indicates the time difference (in hours) between the time of observation and the mid-point of the eclipse. In addition, an illustration of the eclipse phase is shown in the corner of each panel. The illustration shows the line-of-sight component of the rotation velocity of the exposed portion of the stellar photosphere, according to the best-fitting model. Blue indicates a negative (approaching) velocity and red indicates a positive (receding) velocity. The angle is 72° between the sky projections of the primary spin axis and the orbital axis. During the complete phase of the eclipse, much of the approaching portion of the stellar surface is covered, while only a small fraction of the receding portion is blocked. As a result, the primary absorption lines are missing most of their blue wings throughout the eclipse. Had the axes been well-aligned, then in the first half of the eclipse the approaching part of the stellar surface would have been covered, and in the second half of the eclipse the receding part of the stellar surface would have been covered.

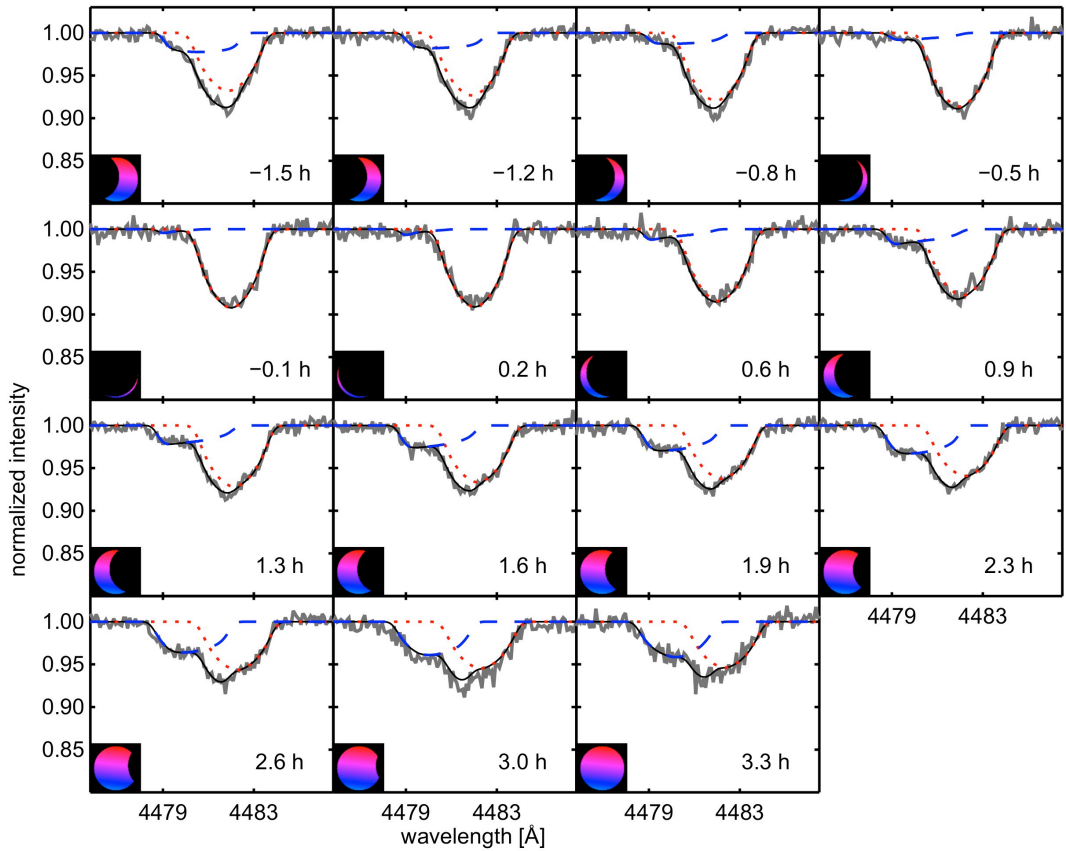

**Supplementary Figure 3 | Observations of DI Herculis during the secondary eclipse.** The same plotting conventions apply as in the previous figure. In this case our best-fitting model has an angle of 84° between the sky projections of the secondary spin axis and the orbital axis. As depicted by the insets, during the secondary eclipse, the approaching and the receding areas of the secondary star's surface are both partly covered by the primary star, to nearly the same extent. This shows that the sky-projected spin axis of the secondary star lies nearly within the orbital plane. Due to the orbital inclination of 89.3°, somewhat more redshifted light is blocked from view.